# Algorithmic Pricing via Virtual Valuations

Shuchi Chawla[*]    Jason D. Hartline[†]    Robert D. Kleinberg[‡]

**Abstract**

*Algorithmic pricing* is the computational problem that sellers (e.g., in supermarkets) face when trying to set prices for their items to maximize their profit in the presence of a known demand. Guruswami et al. (2005) propose this problem and give logarithmic approximations (in the number of consumers) for the unit-demand and single-parameter cases where there is a specific set of consumers and their valuations for bundles are known precisely. Subsequently several versions of the problem have been shown to have poly-logarithmic inapproximability. This problem has direct ties to the important open question of better understanding the Bayesian optimal mechanism in multi-parameter agent settings; however, for this purpose approximation factors logarithmic in the number of agents are inadequate. It is therefore of vital interest to consider special cases where constant approximations are possible.

We consider the unit-demand variant of this pricing problem. Here a consumer has a valuation for each different item and their value for a set of items is simply the maximum value they have for any item in the set. Instead of considering a set of consumers with precisely known preferences, like the prior algorithmic pricing literature, we assume that the preferences of the consumers are drawn from a distribution. This is the standard assumption in economics; furthermore, the setting of a specific set of customers with specific preferences, which is employed in all of the prior work in algorithmic pricing, is a special case of this general Bayesian pricing problem, where there is a discrete Bayesian distribution for preferences specified by picking one consumer uniformly from the given set of consumers. Notice that the distribution over the valuations for the individual items that this generates is obviously correlated. Our work complements these existing works by considering the case where the consumer's valuations for the different items are independent random variables. Our main result is a constant approximation algorithm for this problem that makes use of an interesting connection between this problem and the concept of *virtual valuations* from the single-parameter Bayesian optimal mechanism design literature.

## 1 Introduction

It is vital to the study of resource allocation in distributed settings such as the Internet that inherent economic issues be addressed. Recently the area of *algorithmic pricing* (Aggarwal et al., 2004; Balcan and Blum, 2006; Briest and Krysta, 2007; Guruswami et al., 2005; Hartline and Koltun, 2005) has emerged as a setting for studying optimization in resource allocation in the presence of a natural fairness constraint: there is a uniform pricing rule under which the consumers are allowed to choose the allocation they most desire. This area has important connections to *algorithmic mechanism design* (Balcan et al., 2005; Guruswami et al., 2005; Aggarwal and Hartline, 2006) in addition to obvious applications in traditional market settings such as pricing items in a supermarket.

A *pricing* can be thought of as a menu listing the prices for all possible allocations to a consumer. Given a pricing, a consumer's *preference* indicates a most desired allocation. The algorithmic pricing problem, then, is to take an instance given by a class of allowable pricings and a set of consumers, and compute the pricing maximizing (or approximately maximizing) a specific objective. An *item-pricing* is one where each individual item is assigned a price, and the price of any bundle is the sum of the prices of the items in the bundle. In this arena, the objective of maximizing the profit of the seller presents significant challenges, even when there are no supply constraints. Indeed when the items are pure complements, i.e., consumers are single-minded and combinatorial; and when the items are pure substitutes, i.e., consumers have unit demands, recent works show hardness results for item-pricing that essentially match the poor performance of trivial algorithms (Demaine et al., 2006; Briest and Krysta, 2007). This motivates the search for relevant special cases where algorithmic theory gives an improved understanding of pricing.

[*]Computer Sciences Department, University of Wisconsin - Madison. Email: shuchi@cs.wisc.edu.
[†]Electrical Engineering and Computer Science, Northwestern University. Email: hartline@eecs.northwestern.edu.
[‡]Dept. of Computer Science, Cornell University. Email: rdk@cs.cornell.edu.

Algorithmic pricing and algorithmic mechanism design have important connections. Indeed for unlimited supply profit maximization in prior-free settings (that is, when the seller has no prior information about the consumers' preferences), Balcan et al. (2005) give a general reduction from truthful mechanism design to algorithmic pricing. These results are important, in particular, as they address the challenging open problem of optimal mechanism design in multi-parameter settings (e.g., general combinatorial auctions and multi-item unit-demand auctions).

Optimal mechanism design is well-understood in single parameter Bayesian settings (e.g., a single-item auction)—the well known result of Myerson (1981) gives a closed form characterization for the optimal mechanism. No such characterization is known in general for optimal mechanisms in multi-parameter settings (e.g. where consumers in interested in multiple different items or services). In particular, while the single-parameter optimization problem satisfies several nice properties, e.g., the optimal mechanism is always deterministic, these fail to carry over to the multi-parameter setting. Characterizing the optimal solution to the multi-parameter mechanism design problem in special cases has been a key area of focus in mechanism design over the past few decades. In the special case where there is a single customer, finding the optimal deterministic mechanism is equivalent to finding the optimal pricing. Unfortunately as for mechanisms, optimal pricings do not admit simple closed-form characterizations in general. Armstrong (2006) and Stole (2007) survey some recent results in the economics literature on pricing in special cases.

In this paper we bring these results full-circle by showing that techniques from Myerson's optimal auction in the single-parameter setting are useful for the unit-demand pricing problem, and allow us to obtain a simple characterization as well as a polynomial time algorithm for *near-optimal* pricings in this setting. By designing a pricing to mimic the reserve prices of an optimal auction, we are able to approximate the profit of the optimal auction (and thus, also, the profit of the optimal pricing). Of course, a key difference between single-item auctions and unit-demand pricing is that in an auction the different bids compete against each other, while in a pricing problem there is no competition. Accordingly, our pricing algorithm "simulates competition" by setting higher reserve values.

Formally, the algorithmic mechanism design and algorithmic pricing problems are defined as follows:

**Definition 1 (Bayesian Single-item Auction Problem (BSAP))**
*Given,*
- *a single item for sale,*
- *$n$ consumers, and*
- *distribution $\mathbf{F}$ from which consumer valuations are drawn.*

*Goal: design seller optimal auction for $\mathbf{F}$.*

**Definition 2 (Bayesian Unit-demand Pricing Problem (BUPP))**
*Given,*
- *a single unit-demand consumer,*
- *$n$ items for sale, and*
- *distribution $\mathbf{F}$ from which the consumer's valuations for each item are drawn.*

*Goal: compute seller optimal item-pricing for $\mathbf{F}$.*

In the special case where $\mathbf{F}$ is the product distribution $F_1 \times \cdots \times F_n$, the *Bayesian single-item auction problem* was solved by Myerson in his seminal paper on mechanism design (Myerson, 1981). His solution is based on determining the allocation of the item for sale using the consumers' *virtual valuations* (see Section 2), instead of their actual valuations. We consider product distributions for the *Bayesian unit-demand pricing problem* and show that:

- The optimal revenue of a single-item auction is an upper bound on the revenue of the optimal unit-demand pricing.

- The optimal unit-demand pricing that uses a single *virtual price*[1] for all items obtains a constant fraction of the revenue of the optimal auction.

- If all the distributions satisfy the monotone hazard rate condition (defined in Section 2), a nearly-optimal virtual price can be computed in polynomial time.

We first demonstrate the connection between BSAP and BUPP in the context of identically distributed valuations, i.e. when $F_i = F_j$ for all $i \ne j$. In this i.i.d. case, our algorithm outputs the same price (not just the same virtual price) for all the items. Note that it is easy to optimize revenue over the space of all pricings that price each item at the same value—just consider the distribution of the maximum valuation ($\max_i v_i$) and solve this problem as a single-consumer single-item revenue maximization problem. One might expect that in the i.i.d. case this optimal single price is in fact the overall optimal pricing. However, a simple example shows that this is not true—consider two items, each with a value independently equal to 1 with probability 2/3 and 2 with probability 1/3; then a simple calculation shows that the pricings $(1, 2)$ and $(2, 1)$ are optimal with respect to revenue[2] and the pricings $(1, 1)$ and $(2, 2)$ are strictly sub-optimal. In Section 3.2 we prove that in the i.i.d. case, the revenue of the optimal single-price solution is a 2.17-approximation to the optimal revenue of the single-item auction.

We extend this result to the case of general product distributions in Sections 3.1 and 4, proving that the revenue of the optimal single virtual-price solution is a 3-approximation to the optimal revenue of the corresponding BSAP.

We first develop our main argument for distributions that satisfy the monotone hazard rate condition and more generally for so-called *regular* distributions (see Definitions 3 and 4 in Section 2) in Section 3.1. In Section 4 we consider distributions that are not regular, and in particular do not satisfy the monotone hazard rate condition. Myerson's solution to BSAP in this case uses a smoothed or "ironed" version of virtual valuations. We show that the same fix can be applied to the pricing problem and again the revenue of the optimal single *ironed-virtual-price* solution is a 3-approximation to the optimal revenue of the single-item auction. This, however, requires extra technical machinery because of the fact that the inverse of an ironed virtual valuation function is not continuous.

We consider the question of computing the optimal virtual price in Section 5. The main challenge here is in inverting the virtual valuation functions. For discrete regular distributions this is straightforward and we obtain a polynomial time 3-approximation. For continuous regular distributions, we consider a computational model in which we have oracle access to the cumulative distribution function and probability density functions the distributions (see Section 2 for more detail). In this case we can compute the inverses of virtual valuations only approximately. Once again we obtain a polynomial-time approximation — our algorithm guarantees a $3 + \epsilon$ approximation with a $1 - o(1)$ probability. A downside of this algorithm is that it relies on sampling a large number of valuation vectors from the given distribution, and is therefore slow. We complement this with a simpler and faster algorithm that guarantees a $6 + \epsilon$ approximation with probability 1. The latter algorithm uses a Vickrey auction as its basis, instead of Myerson's optimal auction.

We leave open the problem of designing a polynomial time approximation algorithm for the non-regular case. The challenge in this case is to come up with a polynomial time algorithm for computing ironed-virtual-valuations.

## 2  Notation and preliminaries

In the problems we consider, the input is a distribution over $n$-tuples of valuations. We use $\mathbf{v} = (v_1, \ldots, v_n)$ for the random variable representing the valuation vector as well as its instantiation. In BUPP we interpret $v_i$ as the valuation of the (single) consumer for item $i$, while in BSAP we interpret $v_i$ as the valuation of consumer $i$ for the (single) item. The value $v_i$ is drawn independently from the distribution $F_i$ over the range $[\ell_i, h_i]$. Following standard notation, we use $\mathbf{v}_{-i}$ to denote all the valuations except the $i$th one. $\mathbf{F} = F_1 \times \cdots \times F_n$ denotes the product distribution from which $\mathbf{v}$ is drawn, and $f_i(v_i)$ denotes the probability density of valuation $v_i$. (Unless otherwise specified, we will assume that the distribution of $v_i$ is absolutely continuous with respect to Lebesgue measure, so the density function $f_i(v_i)$ is well-defined.)

---

[1]The virtual price is a price in virtual valuation space instead of valuation space.

[2]We assume that whenever the consumer faces a tie, i.e. two or more items bring equal utility to her, the seller has the ability to break the tie in favor of any of the items (in particular, the most expensive item). The seller could enforce this by giving a negligibly small discount to the consumer for the most expensive item.

### Regularity and the Monotone Hazard Rate condition

In much of the paper we will assume that the distributions $F_i$ satisfy the so-called *regularity* condition defined below. This condition is satisfied by any distribution that has a *monotone hazard rate* (MHR) and is a standard assumption used in economics. In the single-consumer, single-item case, this condition essentially implies that the revenue as a function of price has a unique maximum.

**Definition 3 (Monotone Hazard Rate)** *A one-dimensional distribution $F$ with density $f$ is said to satisfy the* monotone hazard rate *(MHR) condition if the "hazard rate" of the distribution, $\frac{f(v)}{1-F(v)}$, is a monotonically non-decreasing function of $v$.*

**Definition 4 (Regularity)** *A one-dimensional distribution $F$ with density $f$ is said to be* regular *(or satisfy regularity) if $v - \frac{1-F(v)}{f(v)}$ is monotonically non-decreasing for all $v$.*

When each of the $F_i$s satisfy regularity, we say that the product distribution $\mathbf{F} = F_1 \times \cdots \times F_n$ is regular. In Section 4 we extend our results to distributions that do not satisfy regularity. This is also called the non-regular case in the literature.

### Virtual valuations and the Bayesian Single-item Auction Problem (BSAP)

The Bayesian Single-item Auction Problem (BSAP) is described as follows: there is a single item for sale and $n$ bidders with values given by the vector $\mathbf{v}$; each bidder's value $v_i$ is drawn independently from a distribution $F_i$; the goal of the mechanism designer is to design an *incentive-compatible* auction so as to maximize the revenue obtained by the seller from the sale of the item.

In one of the seminal works of Bayesian mechanism design, Myerson developed a mechanism for this problem that obtains the maximum revenue for the seller over the class of all incentive-compatible mechanisms (Myerson, 1981). Myerson's mechanism (denoted $\mathcal{M}$ hereafter) is based on the following definitions:

**Definition 5** *The* virtual valuation *of bidder $i$ with valuation $v_i$ drawn from $F_i$ is*

$$\phi_i(v_i) = v_i - \frac{1 - F_i(v_i)}{f_i(v_i)}. \tag{1}$$

*The* virtual surplus *of a single-item auction is the virtual valuation of the winner.*

Note that for any regular distribution, the virtual valuation $\phi(v)$ is a non-decreasing function of the valuation $v$.

**Theorem 1 (Myerson (1981))** *Any incentive-compatible auction $\mathcal{A}$ has expected revenue equal to its expected virtual surplus.*

The virtual valuation of a bidder essentially denotes the marginal revenue obtained by allocating the item to this bidder. A simple consequence of Myerson's Theorem is that maximizing revenue (in expectation) is equivalent to maximizing virtual surplus; therefore, for regular distributions the optimal single-item auction is the one that sells to the bidder with the highest non-negative virtual valuation[3]. Incentive compatibility constraints then imply that the payment of the optimal mechanism should be the value at which the winning bidder's virtual valuation equals the second highest virtual valuation, i.e., the payment is equal to the virtual-valuation-inverse of the second highest virtual valuation. (If all other virtual valuations are negative, we consider the second highest virtual valuation to be 0.) It is sometimes convenient to view Myerson's mechanism as offering each bidder a take-it-or-leave-it price, where the price offered to bidder $i$ is equal to $\phi_i^{-1}(\nu_i)$ and $\nu_i = \max_{j \neq i} \max(\phi_j(v_j), 0)$. Notice that only the bidder with the highest virtual valuation would accept such a take-it-or-leave-it offer.

---
[3] In the non-regular case, that is when Definition 4 is not satisfied, this auction may not be incentive compatible. In Section 4 we describe the modifications required to obtain an optimal incentive compatible auction in that case.

We use $\mathcal{R}^\mathcal{A}$ to denote the expected revenue of an incentive-compatible auction $\mathcal{A}$ for BSAP. $\mathcal{R}^\mathcal{M}$ denotes the expected revenue of Myerson's auction $\mathcal{M}$.

**Corollary 2** *When $\mathbf{F}$ satisfies regularity, $\mathcal{R}^\mathcal{M} \geq \mathcal{R}^\mathcal{A}$ for all incentive-compatible auctions $\mathcal{A}$.*

We can also apply Myerson's Theorem to a variant of the single-item auction problem where the seller has some reservation value, $\nu$, for keeping the item. The virtual surplus in this setting would be the virtual valuation of the winning bidder or $\nu$ if the item remains unsold. Myerson's Theorem says that the optimal mechanism (by maximizing virtual surplus) sells the item to the bidder with the highest virtual valuation if and only if that virtual valuation is at least $\nu$. We denote this mechanism by $\mathcal{M}_\nu$.

In other words, if we use the notation $\chi(\mathcal{A})$ to denote the probability that the item is unsold when using a given auction mechanism $\mathcal{A}$, then for every truthful mechanism $\mathcal{A}$ we have:

**Corollary 3** *When $\mathbf{F}$ satisfies regularity, for all incentive-compatible auctions $\mathcal{A}$,*

$$\mathcal{R}^{\mathcal{M}_\nu} + \nu \cdot \chi(\mathcal{M}_\nu) \geq \mathcal{R}^\mathcal{A} + \nu \cdot \chi(\mathcal{A}).$$

The observation of the following fact completes our preliminary discussion of virtual valuations and optimal auctions.

**Fact 4** *Virtual valuations satisfy $\phi_i(v_i) \leq v_i$.*

**The Bayesian Unit-demand Pricing Problem (BUPP)**

The Bayesian Unit-demand Pricing Problem (BUPP) is described as follows: there are $n$ items for sale and a single consumer with unit-demand, quasi-linear preferences given by the vector $\mathbf{v}$; the consumer's value $v_i$ for item $i$ is drawn independently from a distribution $F_i$; the goal of our algorithm is to determine a price vector $\mathbf{p} = (p_1, \cdots, p_n)$ such that the expected revenue $\mathcal{R}^\mathbf{p}$, as defined below, is maximized.

$$\mathcal{R}^\mathbf{p} = \sum_i p_i \cdot \mathbf{Pr}_{\mathbf{v} \sim \mathbf{F}}\left[(v_i - p_i) = \max_{j \leq n}(v_j - p_j)\right]$$

We now give a number of definitions that will allow us to talk about the outcome and performance of a pricing, $\mathbf{p}$.

- (For implicit $\mathbf{p}$) $q_i$ is the probability that $v_i \geq p_i$ when $v_i \sim F_i$. In other words, $q_i = 1 - F_i(p_i)$.

- $\chi(\mathbf{p})$ is the probability no item is sold at pricing $\mathbf{p}$. In other words, $\chi(\mathbf{p}) = \mathbf{Pr}[v_i < p_i, \text{ for all } i]$. Clearly, $\chi(\mathbf{p}) = \prod_i F_i(p_i) = \prod_i (1 - q_i)$.

- The *reserve price function*,[4] $r_i(\cdot)$, is equal to the *inverse virtual valuation function*, $\phi_i^{-1}(\cdot)$.

- $\mathbf{r}(\nu) = (r_1(\nu), \ldots, r_n(\nu))$ is the price vector with *constant virtual price* $\nu$.

- $\nu_x$ is the virtual valuation that satisfies $\chi(\mathbf{r}(\nu_x)) = x$.

---
[4]This terminology comes from single-item settings where a seller's optimal reserve price for an item of value $\nu$ is $\phi^{-1}(\nu)$.

**The connection between BUPP and BSAP**

The connection between BUPP and BSAP starts with the case where $n = 1$ and the two problems are identical. Here the optimal pricing (and the optimal auction) is to offer the item at a price equal to the consumer's inverse virtual valuation of zero, $p_1 = \phi_i^{-1}(0)$.

For larger $n$, the competition between bidders in BSAP will allow the optimal auction, $\mathcal{R}^{\mathcal{M}}$, to obtain at least the revenue, $\mathcal{R}^{\mathbf{p}}$, of any pricing $\mathbf{p}$ for BUPP.

**Lemma 5** *For any price vector* $\mathbf{p}$, $\mathcal{R}^{\mathcal{M}} \geq \mathcal{R}^{\mathbf{p}}$.

*Proof:* For a given pricing $\mathbf{p}$, consider the following mechanism $\mathcal{A}_{\mathbf{p}}$: given a valuation vector $\mathbf{v}$, we allocate the item to the bidder $i$ with $v_i \geq p_i$ that maximizes $v_i - p_i$. Prices are determined by the standard "threshold payment" rule: the winning bidder, $i$, pays the minimum bid value which would still make $i$ the winner. $\mathcal{A}_{\mathbf{p}}$ is truthful because it gives a monotone allocation procedure: if a winning bidder unilaterally increases her bid, she still wins. Therefore, $\mathcal{R}^{\mathcal{A}_{\mathbf{p}}} \leq \mathcal{R}^{\mathcal{M}}$.

Now consider any valuation vector $\mathbf{v}$ and suppose that $\mathcal{A}_{\mathbf{p}}$ allocates the item to bidder $i$. Then the minimum bid at which this bidder is allocated the item is $p_i + \max_{j \neq i}(v_j - p_j, 0)$, which is at least $p_i$. Therefore, the revenue of $\mathcal{A}_{\mathbf{p}}$ when the valuation vector is $\mathbf{v}$ is at least $p_i$. However, the revenue of the pricing $\mathbf{p}$ when the valuation vector is $\mathbf{v}$ is exactly $p_i$. Therefore, $\mathcal{R}^{\mathcal{A}_{\mathbf{p}}} \geq \mathcal{R}^{\mathbf{p}}$. Combining the two inequalities proves the lemma. □

Our motivation to relate BUPP to Myerson's solution of BSAP stems from the observation that as the number of bidders gets large (especially in the case of identically distributed valuations), the price offered to a bidder in Myerson's mechanism becomes tightly concentrated around a single value (the expectation of the virtual-value-inverse of the maximum over virtual valuations of other bidders). This value is a reasonable candidate for the price of item $i$ in the pricing problem, and is indeed roughly what we use (with some modifications to allow for a simpler analysis). Thus, in examining the actual outcomes of the optimal auction, we gain intuition for outcomes our pricing should attempt to mimic.

This approach has obstacles that must be overcome. Myerson's mechanism, by allowing the price to be an appropriate function of other bidders' values, ensures that only one bidder accepts the offered price; whereas in BUPP, with some probability, more than one of the values is above its corresponding offer price. When this happens the consumer gets to pick which item to buy. The price earned by our solution to BUPP may be much worse than the price earned by Myerson's solution to BSAP for the same valuation vector. We can get around this problem by making use of regularity and the monotone hazard rate assumption (and other techniques from optimal mechanism design when regularity does not hold).

**The computational model**

Much of this paper focuses on an analysis that reduces the multi-dimensional price optimization problem to a single-dimensional uniform virtual price optimization. To actually compute (or approximate) the pricing that our analysis suggests, we consider two different computational models:

- **(Discrete explicit)** In this model, each of the distributions $F_i$ is a discrete distribution with small support. These distributions are specified explicitly, and our algorithm is required to run in time polynomial in the number of items $n$, and the size of the largest support.

- **(Continuous with oracles)** In this model, the distributions $F_i$ are continuous with known supports $[\ell_i, h_i]$. The algorithm is provided the following oracles: an oracle to determine $F_i(v)$ given a value $v$ and an index $i$, an oracle to determine the density $f_i(v)$ given a value $v$ and an index $i$, an oracle to sample from the product distribution $\mathbf{F}$, and finally an oracle for $r_i(0) = \phi_i^{-1}(0)$ for all $i$.[5] The algorithm is required to run in time polynomial in the number of items $n$ and the range $(\max_i h_i)/(\min_i \ell_i)$.

---

[5] We can remove the last assumption about $r_i(0)$ being given for all $i$ using the techniques we give. However, it is natural to assume that they are known as they are necessary for the solution of the "solved" problem of optimally pricing a single item.

# 3 Approximating pricing in the regular case

In this section we demonstrate that the unit-demand optimal pricing that uses a single virtual price for all items obtains a constant fraction of the revenue of the optimal single-item auction (and also of the revenue of the optimal pricing). Specifically we obtain a 3-approximation for general distributions, and an improved 2.17-approximation when all the valuations are distributed identically.

## 3.1 Analysis of the general case

In this section we show that for $\nu = \max(0, \nu_{1/2})$ the pricing $\mathbf{p} = \mathbf{r}(\nu)$ satisfies

$$\mathcal{R}^{\mathbf{p}} \geq \mathcal{R}^{\mathcal{M}}/3.$$

That is, a single price in virtual valuation space approximates the optimal auction (and thus, by Lemma 5, the optimal pricing). We first show that $\mathcal{R}^{\mathcal{M}}$ can be bounded from above in terms of $\nu$ and $\mathcal{R}^{\mathcal{M}_\nu}$ (below, Corollary 6). We then bound from above both $\nu$ (below, Lemma 7) and $\mathcal{R}^{\mathcal{M}_\nu}$ (below, Lemmas 8, 9, Corollaries 10, 11) by appropriate multiples of $\mathcal{R}^{\mathbf{p}}$.

**Corollary 6** *For any $\nu \geq 0$, $\mathcal{R}^{\mathcal{M}_\nu} + \nu \cdot \chi(\mathcal{M}_\nu) \geq \mathcal{R}^{\mathcal{M}}$.*

*Proof:* This result follows by invoking Corollary 3 with $\mathcal{A} = \mathcal{M}$ and noting that $\chi(\mathcal{A}) \geq 0$ for every mechanism $\mathcal{A}$. □

**Lemma 7** *For $\mathbf{p} = \mathbf{r}(\nu)$, $\mathcal{R}^{\mathbf{p}} \geq (1 - \chi(\mathbf{p})) \cdot \nu$.*

*Proof:* Fact 4 implies that $p_i = \phi_i^{-1}(\nu) \geq \nu$ for all $i$. By definition, $1 - \chi(\mathbf{p})$ is the probability that an item is sold. Thus, $\mathcal{R}^{\mathbf{p}} \geq (1 - \chi(\mathbf{p})) \cdot \nu$. □

**Lemma 8** *For any $\mathbf{p}$, $\mathcal{R}^{\mathbf{p}} \geq \chi(\mathbf{p}) \cdot \sum_i p_i q_i$.*

*Proof:* The revenue $\mathcal{R}^{\mathbf{p}}$ is bounded below by the summation, over all $i$, of $p_i$ times the probability that $i$ is the unique index satisfying $v_i \geq p_i$, i.e.

$$\begin{aligned}\mathcal{R}^{\mathbf{p}} &\geq \sum_i p_i \cdot \left( q_i \prod_{j \neq i} (1 - q_j) \right) \\ &= \sum_i p_i q_i \tfrac{\chi(\mathbf{p})}{1-q_i} \geq \chi(\mathbf{p}) \sum_i p_i q_i.\end{aligned}$$

□

Before we bound $\mathcal{R}^{\mathcal{M}_\nu}$ in terms of $\mathcal{R}^{\mathbf{p}}$ we need a new definition. For a given auction $\mathcal{A}$ and prices $\mathbf{p}$, consider the event, $\mathcal{E}_{\mathcal{A},\mathbf{p}}$, in which the winner, say $i$, has $v_i \geq p_i$. Let $\mathcal{R}^{\mathcal{A}}_{\mathbf{p}}$ be the contribution to the expected revenue of $\mathcal{A}$ from such events. That is, $\mathcal{R}^{\mathcal{A}}_{\mathbf{p}}$ is the expected revenue of $\mathcal{A}$ conditioned on $\mathcal{E}_{\mathcal{A},\mathbf{p}}$ times the probability that $\mathcal{E}_{\mathcal{A},\mathbf{p}}$ happens. The following lemma shows that $\mathcal{R}^{\mathcal{A}}_{\mathbf{p}}$ can be bounded in terms of $\mathcal{R}^{\mathbf{p}}$.

**Lemma 9** *Under regularity, for any $\mathbf{p} \geq \mathbf{r}(0)$ and any incentive-compatible auction, $\mathcal{A}$, we have $\mathcal{R}^{\mathcal{A}}_{\mathbf{p}} \leq \sum_i p_i q_i$.*

*Proof:* We modify $\mathcal{A}$ to get $\mathcal{A}'$ with $\mathcal{R}^{\mathcal{A}}_{\mathbf{p}} \leq \mathcal{R}^{\mathcal{A}'}_{\mathbf{p}} = \mathcal{R}^{\mathcal{A}'}$. If $\mathcal{A}$ sells to a bidder $i$ with valuation at least $p_i$, $\mathcal{A}'$ does the same; otherwise, $\mathcal{A}'$ does not sell the item[6]. It is easy to see that $\mathcal{A}'$ is incentive compatible if $\mathcal{A}$ is. Since the modified auction never sells to a bidder $i$ with $v_i < p_i$, it is immediate that $\mathcal{R}^{\mathcal{A}'}_{\mathbf{p}} = \mathcal{R}^{\mathcal{A}'}$. Moreover, if the winner $i$ of $\mathcal{A}$ has value $v_i \geq p_i$, this modification does not change the allocation and only increases the payments. So $\mathcal{R}^{\mathcal{A}}_{\mathbf{p}} \leq \mathcal{R}^{\mathcal{A}'}_{\mathbf{p}}$. Now we get an upper bound on $\mathcal{R}^{\mathcal{A}'}$.

---

[6]Notice that incentive compatibility constraints require that the payment in $\mathcal{A}'$ when allocating to bidder $i$ is maximum of the payment $i$ would have made in $\mathcal{A}$ and $p_i$.

$\mathcal{A}'$ is an auction that never sells to a bidder $i$ at price less than $p_i$. Certainly, its revenue is less than the optimal auction satisfying the same price constraint. Moreover, this optimal auction's revenue is less than the optimal auction that does not have a supply constraint (i.e., it can sell multiple copies of the item). We now show that the optimal auction, that can potentially sell an item to all bidders at once, but is constrained to use prices at least $\mathbf{p}$, has revenue precisely $\sum_i p_i q_i$.

Since it can sell to all bidders, an optimal auction would make its allocation decisions for each bidder independently. Consider bidder $i$. If $v_i \geq p_i$ then $i$ has non-negative virtual surplus. Regularity and Myerson's theorem (expected revenue equals expected virtual surplus) imply that this optimal auction would sell to bidder $i$. Of course when $v_i < p_i$ the auction cannot sell to bidder $i$. The auction that sells to $i$ if and only if $v_i \geq p_i$ has expected payment precisely $p_i q_i$. Summing over all bidders, the total expected revenue of this optimal auction would be $\sum_i p_i q_i$. $\square$

We can now put lemmas 8 and 9 together to get the following:

**Corollary 10** *Under regularity, any auction $\mathcal{A}$ and any pricing $\mathbf{p} \geq \mathbf{r}(0)$ satisfies*

$$\chi(\mathbf{p}) \cdot \mathcal{R}^{\mathcal{A}}_{\mathbf{p}} \leq \mathcal{R}^{\mathbf{p}}.$$

Since $\mathcal{R}^{\mathcal{M}_\nu} = \mathcal{R}^{\mathcal{M}_\nu}_{\mathbf{p}}$ for $\mathbf{p} = \mathbf{r}(\nu)$ we have:

**Corollary 11** *Under regularity, for any $\nu \geq 0$, $\mathbf{p} = \mathbf{r}(\nu)$ satisfies $\chi(\mathbf{p}) \cdot \mathcal{R}^{\mathcal{M}_\nu} \leq \mathcal{R}^{\mathbf{p}}$.*

**Lemma 12** *Under regularity, with $\nu_{1/2} \geq 0$, $\mathbf{p} = \mathbf{r}(\nu_{1/2})$ satisfies $\mathcal{R}^{\mathcal{M}} \leq 3\mathcal{R}^{\mathbf{p}}$.*

*Proof:* Notice that the probability that no item is sold in $\mathcal{M}_{\nu_{1/2}}$ and under pricing $\mathbf{p}$ is the same, i.e., $\chi(\mathcal{M}_{\nu_{1/2}}) = \chi(\mathbf{p}) = 1/2$. Call this probability $x$.

$$\begin{aligned}
\mathcal{R}^{\mathcal{M}} &\leq \mathcal{R}^{\mathcal{M}_{\nu_x}} + \nu_x x & \text{Corollary 6} \\
&\leq \tfrac{1}{x}\mathcal{R}^{\mathbf{p}} + \nu_x x & \text{Corollary 11} \\
&\leq \tfrac{1}{x}\mathcal{R}^{\mathbf{p}} + \tfrac{x}{1-x}\mathcal{R}^{\mathbf{p}} & \text{Lemma 7} \\
&= 3\mathcal{R}^{\mathbf{p}} & \text{Using } x = 1/2.
\end{aligned}$$

Notice that $x = 1/2$ is indeed the optimal choice for $x$, above. $\square$

The only loose end to wrap up now is the case that $\nu_{1/2} < 0$. In this case, we can show that $\mathbf{p} = \mathbf{r}(0)$ is a 2-approximation using only Corollary 10.

**Lemma 13** *Under regularity, with $\nu_{1/2} \leq 0$, $\mathbf{p} = \mathbf{r}(0)$ satisfies $\mathcal{R}^{\mathcal{M}} \leq 2\mathcal{R}^{\mathbf{p}}$.*

*Proof:* For such a pricing, $\chi(\mathbf{p}) > 1/2$ and $\mathcal{R}^{\mathcal{M}}_{\mathbf{p}} = \mathcal{R}^{\mathcal{M}}$. Thus, Corollary 10 applied to $\mathcal{M}$ shows that $2\mathcal{R}^{\mathbf{p}} \geq \mathcal{R}^{\mathcal{M}}$. $\square$

We combine Lemmas 5, 12, and 13 to get the main theorem of the paper.

**Theorem 14** *Under regularity and with $\nu = \max(0, \nu_{1/2})$, the pricing $\mathbf{p} = \mathbf{r}(\nu)$ is a 3-approximation to the optimal pricing.*

## 3.2 Analysis of the i.i.d. case

We now consider the i.i.d. case where all valuations are distributed according to $F$, with density function $f$, and virtual valuation function $\phi(\cdot)$. In this case, our solution of the form $\mathbf{r}(\nu)$ for some $\nu$ is a single-value pricing, that is, it charges the same price for every item. Constrained to this class of single-value pricings, it is easy to analytically describe the optimal pricing and even easier to compute it via a sampling algorithm. Let $F_{\max}$ be the distribution for $\max_i v_i$ when $v_i$ is distributed from $F_i$. The optimal single price to use is the $p$ that maximizes $p \cdot (1 - F_{\max}(p))$.

Recall that we showed in the introduction that a single-value pricing is not necessarily optimal. In this section we show that such a pricing is in fact fairly close to an optimal pricing in the i.i.d. case. For a pricing **p** that is implicit, we let $q$ be the probability that a valuation drawn from $F$ is at least $p$. We will abuse notation to let $\nu_{1/e}$ be the virtual valuation for which $q = 1/n$, i.e., $\nu_{1/e}$ satisfies $1 - F(\phi^{-1}(\nu_{1/e})) = 1/n$. As before we choose $\nu = \max(0, \nu_{1/e})$ and consider the pricing $\mathbf{p} = \mathbf{r}(\nu)$.

We show that this pricing is a 2.17-approximation to the optimal pricing. Our analysis is near tight— towards the end of this section we give an example for which no single-value pricing is better than a $(2-o(n))$-approximation to the optimal pricing; Therefore, to beat the factor of 2, one must necessarily consider non-single-value pricings.

The motivation for our abuse of notation is the following lemma.

**Lemma 15** *For distribution $F$ and pricing $\mathbf{p} = (p, \ldots, p)$ such that $1 - F(p) = 1/n$, we have $\chi(\mathbf{p}) \leq 1/e$, and this bound is asymptotically tight.*

*Proof:* By definition, $\chi(\mathbf{p}) = (F(p))^n = (1 - 1/n)^n \leq 1/e$. □

The following adaptations of results from the preceding section allow us to prove the main result of this section. The improvement we obtain in the i.i.d. case, over the general case, comes from the following. First, Fact 16 gives a slightly better lower bound on $\mathcal{R}^\mathbf{p}$ than Lemma 7. Second, we can use this bound (instead of Lemma 8) with Lemma 9, to obtain an improved upper bound on $\mathcal{R}^{\mathcal{M}_\nu}$ in terms of $\mathcal{R}^\mathbf{p}$ when $\nu_{1/e} \geq 0$ and $\chi(\mathbf{p}) < 1/e$.

**Fact 16** $\mathcal{R}^\mathbf{p} = p(1 - \chi(\mathbf{p}))$ *for any* $\mathbf{p} = (p, \ldots, p)$.

**Lemma 17** *Under regularity, with $\nu_{1/e} \geq 0$, $\mathbf{p} = \mathbf{r}(\nu_{1/e})$ satisfies $\mathcal{R}^\mathcal{M} \leq 2.17 \mathcal{R}^\mathbf{p}$.*

*Proof:* Notice that the probability that no item is sold in $\mathcal{M}_{\nu_{1/e}}$ and under pricing **p** is the same, i.e., $x = \chi(\mathcal{M}_{\nu_{1/e}}) = \chi(\mathbf{p}) \leq 1/e$.

$$\begin{aligned}
\mathcal{R}^\mathcal{M} &\leq \mathcal{R}^{\mathcal{M}_{\nu_{1/e}}} + \nu_{1/e} \cdot x & \text{Lemma 6} \\
&\leq \sum_i pq + \nu_{1/e} \cdot x & \text{Lemma 9} \\
&\leq p + \nu_{1/e} \cdot x & \text{Since } q = 1/n \\
&\leq (1 + x) \cdot p & \text{Since } \nu_{1/e} \leq p \\
&= \tfrac{1+x}{1-x} \mathcal{R}^\mathbf{p} & \text{Fact 16.}
\end{aligned}$$

Finally, for any $0 < x \leq 1/e$, $(1+x)/(1-x)$ is at most 2.17. □

We now need to handle the case where $\nu_{1/e} < 0$ by showing that $\mathbf{p} = \mathbf{r}(0)$ is good. Unfortunately, we only know that $\chi(\mathbf{p})$ is approximately at least $1/e$ whereas Lemma 13 needed $\chi(\mathbf{p}) \geq 1/2$. What is needed is a tighter lower bound on $\mathcal{R}^\mathbf{p}$ in terms of $\sum_i p_i q_i = npq$ than is given by Lemma 8.

**Lemma 18** *Any pricing $\mathbf{p} = (p, \ldots, p)$ with $q = 1 - F(p) \leq 1/n$ satisfies $\mathcal{R}^\mathbf{p} \geq npq/2$.*

*Proof:* For the pricing **p**, the probability that a sale is made is exactly

$$\mathbf{Pr}[\exists i \mid v_i \geq p] = 1 - (1-q)^n.$$

Using Taylor's expansion and $q \leq 1/n$ we can simplify this expression as follows: $(1-q)^n < 1 - qn + \tfrac{1}{2}q^2n^2 \leq 1 - \tfrac{1}{2}qn$. Therefore the expected revenue of **p** is $\mathcal{R}^\mathbf{p} = p(1 - (1-q)^n) \geq npq/2$. □

**Lemma 19** *Under regularity, with $\nu_{1/e} \leq 0$, pricing $\mathbf{p} = \mathbf{r}(0)$ satisfies $\mathcal{R}^\mathcal{M} \leq 2\mathcal{R}^\mathbf{p}$.*

*Proof:* First, $q$ is at most $1/n$ so Lemma 18 implies that $\mathcal{R}^\mathbf{p} \geq npq/2$. Myerson's mechanism never sells at price less than $\mathbf{p} = \mathbf{r}(0)$ so Lemma 9 implies that $\mathcal{R}^\mathcal{M} = \mathcal{R}^\mathcal{M}_\mathbf{p} \leq npq$. Thus, $\mathcal{R}^\mathbf{p} \geq \mathcal{R}^\mathcal{M}/2$. □

We combine Lemmas 5, 19, and 17 to get the main theorem of this section.

**Theorem 20** *Under regularity and with $\nu = \max(0, \nu_{1/e})$, the pricing $\mathbf{p} = \mathbf{r}(\nu)$ is a 2.17-approximation to the optimal pricing.*

The following lower bound shows that this theorem is nearly tight for single-value pricings $\mathbf{p} = (p, \ldots, p)$.

**Lemma 21** *There exists a distribution $\mathbf{F} = F_1 \times \cdots \times F_n$, with $F_i = F_j$ for all $i, j$, for which no single-value pricing is better than a $(2 - o(n))$-approximation to the optimal pricing.*

*Proof:* Consider the i.i.d. distribution given by,

$$\mathbf{Pr}[v_i = n] = \tfrac{1}{n^2}$$
$$\mathbf{Pr}[v_i = 1] = 1 - \tfrac{1}{n^2}.$$

The optimal pricing sets $p_1^* = 1$ and $p_i^* = n$ for all $i > 1$. This achieves a revenue of $2 - o(1)$. However, for every pricing $\mathbf{p} = (p, \ldots, p)$ has $\mathcal{R}^{\mathbf{p}} \leq 1$.[7] □

### 3.3 Single-value pricings for general distributions.

Given the simplicity of our solution for i.i.d. distributions, it is natural to ask how well single-value pricings perform for general distributions. The example below shows that for general distributions single-value pricings cannot approximate the optimal pricing to within a constant factor, necessitating a more complicated solution, such as the one we give in Section 3.1.

**Lemma 22** *There exists a distribution $\mathbf{F}$ for which no single-value pricing is a $o(\log n)$-approximation to the optimal pricing.*

*Proof:* Consider the following distribution for $v_i$:

$$\mathbf{Pr}\left[v_i = \tfrac{n}{i}\right] = \tfrac{1}{n}$$
$$\mathbf{Pr}[v_i = 1] = 1 - \tfrac{1}{n}.$$

The pricing $\mathbf{p}$ which sets $p_i = n/i$ achieves a profit of $\Omega(\log n)$ because

$$\begin{aligned}
\mathcal{R}^{\mathbf{p}} &= \sum_{i=1}^{n} \left(\frac{n}{i}\right) \mathbf{Pr}\left[v_i = \frac{n}{i} \text{ and } \forall j < i \ v_j = 1\right] \\
&= \sum_{i=1}^{n} \left(\frac{n}{i}\right) \cdot \left(\frac{1}{n}\right) \cdot \left(1 - \frac{1}{n}\right)^{i-1} \\
&> \sum_{i=1}^{n} \left(\frac{n}{i}\right) \cdot \left(\frac{1}{n}\right) \cdot \left(\frac{1}{4}\right) \\
&= H_n/4,
\end{aligned}$$

where $H_n$ denotes the $n$-th harmonic number, $\sum_{i=1}^{n} \frac{1}{i}$.

On the other hand, if $\mathbf{p}$ is any pricing which sets $p_i = p$ for some fixed value of $p > 1$, then

$$\begin{aligned}
\mathcal{R}^{\mathbf{p}} &\leq \sum_{i=1}^{n} p \cdot \mathbf{Pr}[v_i \geq p] \\
&= \sum_{1 \leq i \leq n/p} p \cdot \left(\frac{1}{n}\right) + \sum_{i > n/p} p \cdot 0 \\
&\leq 1.
\end{aligned}$$

This example establishes that no single-price pricing which can achieve an $o(\log n)$-approximation to the profit of the optimal pricing. □

---

[7]In this example the revenue of Myerson's auction is nearly equal to 2, so the example does not prove any separation between the revenue of Myerson's auction and that of the optimal pricing.

# 4 The non-regular case

In our analysis in Section 3.1, we used the MHR condition to imply that the functions $\phi_i(v_i)$ are non-decreasing, which in turn allowed us to bound the revenue $\mathcal{R}^{\mathcal{A}}_{\mathbf{p}}$ for an auction $\mathcal{A}$ in Lemma 9. When the MHR condition (or regularity) does not hold, Myerson applies a fix to the problem by smoothing out or "ironing" the virtual valuation function to make it a non-decreasing function of $v_i$. We now show that by picking a pricing based on ironed virtual valuations instead of the actual virtual valuations, we achieve exactly the same guarantee as in the regular case—the revenue of our pricing is within a factor of 3 of the revenue of Myerson's mechanism. This result requires new ideas in addition to our approach in Section 3. The main issue we need to deal with is that ironed virtual valuations do not have unique inverses, and we need to pick inverses carefully in order to avoid worsening the approximation factor[8].

We briefly describe this ironing procedure below. The reader is referred to Myerson's paper (Myerson, 1981) and a survey of Bulow and Roberts (1989) for more details.

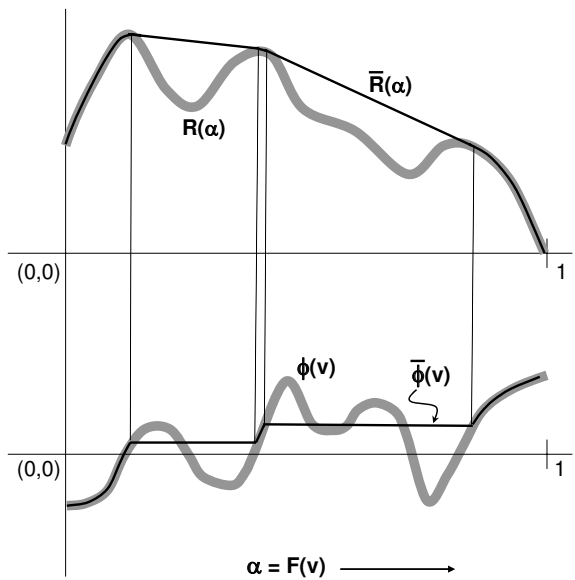

Figure 1: Converting a virtual valuation function $\phi$ to $\bar{R}$ and $\bar{\phi}$

**The ironing procedure**

The ironed virtual valuation function is defined as follows. Consider a single bidder with value $v$ distributed according to function $F$. We assume that the density function $f(v)$ is non-zero for all $v \in [\ell, h]$. For $\alpha \in [0, 1]$, let $R(\alpha)$ denote the revenue generated from offering the item to this bidder at price $F^{-1}(\alpha)$:[9]

$$R(\alpha) = F^{-1}(\alpha)(1-\alpha) = \int_{F^{-1}(\alpha)}^{h} \phi(t) f(t) dt$$

Let $\bar{R}(\alpha)$ be the least-valued concave function on $[0,1]$ with $\bar{R}(\alpha) \geq R(\alpha)$ for all $\alpha$ in that range (see Figure 1). Since $\bar{R}$ is concave, it is differentiable on a dense subset of $[\ell, h]$ (cf. Rockafellar, 1970, Theorem 25.5). Let $\bar{r}(\alpha)$ denote the derivative of $\bar{R}$ wherever defined. The ironed virtual valuation function is defined as below wherever $\bar{r}$ is defined, and is extended to the full range of $v$ by right continuity.

---

[8] A previous version of this paper contained a worse 4-approximation in the non-regular case.
[9] Note that $F^{-1}(\alpha)$ is well-defined because $F$ is a strictly increasing function.

$$\bar{\phi}(v) = -\bar{r}(F(v))$$

Note that since $\bar{R}(\alpha)$ is concave and $F(v)$ is non-decreasing, $\bar{\phi}(v)$ is a non-decreasing function. Furthermore, observing that $\bar{R}(1) = R(1) = 0$, we get the following:

$$\int_{t=v}^{t=h} \bar{\phi}(t) f(t) dt = -\int_{t=F(v)}^{t=1} \bar{r}(t) dt = \bar{R}(F(v)) \tag{2}$$

**Theorem 23 (Myerson (1981))** *The expected revenue of any incentive-compatible auction $\mathcal{A}$ is no more than its expected ironed virtual surplus. Furthermore, if for all $i$, $\mathcal{A}$ has constant probability of allocating to bidder $i$ over any valuation range for which the ironed virtual valuation of bidder $i$ is constant, then the expected revenue of $\mathcal{A}$ is equal to its expected ironed virtual surplus.*

Fortunately, the allocation rule that maximizes ironed virtual surplus (e.g., in BSAP) would naturally have a constant probability of allocating to bidder $i$ over any valuation range for which the ironed virtual valuation of bidder $i$ is constant. Therefore, $\mathcal{M}$ is simply the auction that maximizes ironed virtual surplus. For BSAP, $\mathcal{M}$ first computes the ironed virtual valuations of the values of all bidders. It then allocates the item to the bidder with the highest non-negative ironed virtual valuation at a price equal to the inverse of the second highest one[10]. If there are ties, i.e., two or more bidders with maximal ironed virtual surplus then $\mathcal{M}$ must break this tie constently. (A consistent rule, for example, is to break ties in favor of bidder $i$ over bidder $j$ when $i < j$.)

**Corollary 24** $\mathcal{R}^{\mathcal{M}} \geq \mathcal{R}^{\mathcal{A}}$ *for all incentive-compatible auctions $\mathcal{A}$.*

**Pricings based on ironed virtual values**

Since the ironed virtual valuation functions are monotone but not strictly so, their inverses are not well defined. The standard interpretation of the inverse of an ironed virtual valuation $\nu$ is the infimum of the set of values $v$ such that $\nu = \phi(v)$. However, it will be useful for us to consider both the infimum and the supremum of this set. So, we define $\overleftarrow{\phi}_i^{-1}(\nu) = \inf\{v : \bar{\phi}_i(v) = \nu\}$, and $\overrightarrow{\phi}_i^{-1}(\nu) = \sup\{v : \bar{\phi}_i(v) = \nu\}$.[11]

Let $\nu_x$ be the ironed virtual valuation $\nu$ for which

$$\chi(\overleftarrow{\phi}_1^{-1}(\nu), \overleftarrow{\phi}_2^{-1}(\nu), \cdots, \overleftarrow{\phi}_n^{-1}(\nu)) \leq x$$

and,

$$\chi(\overrightarrow{\phi}_1^{-1}(\nu), \overrightarrow{\phi}_2^{-1}(\nu), \cdots, \overrightarrow{\phi}_n^{-1}(\nu)) \geq x$$

Ideally we would like to find a pricing $\mathbf{p}$ with $\chi(\mathbf{p}) = x$ such that each coordinate $p_i$ is equal to either $\overleftarrow{\phi}_i^{-1}(\nu_x)$ or $\overrightarrow{\phi}_i^{-1}(\nu_x)$. Unfortunately, such a pricing does not always exist. Instead we show below that there exists a vector $\mathbf{p}$ with $\chi(\mathbf{p}) = x$, and *all but one* of the coordinates equal to the corresponding $\overleftarrow{\phi}_i^{-1}(\nu_x)$ or $\overrightarrow{\phi}_i^{-1}(\nu_x)$. We then consider as our solution, one of two pricings obtained by rounding off this last coordinate to $\overleftarrow{\phi}_i^{-1}(\nu_x)$ or $\overrightarrow{\phi}_i^{-1}(\nu_x)$. (Note that these two pricings no longer have their $\chi()$ values equal to $x$.)

Formally, for all $i \in [0, n]$, let $z_i$ denote the pricing

$$(\overrightarrow{\phi}_1^{-1}(\nu_x), \cdots, \overrightarrow{\phi}_i^{-1}(\nu_x), \overleftarrow{\phi}_{i+1}^{-1}(\nu_x), \cdots, \overleftarrow{\phi}_n^{-1}(\nu_x)),$$

---

[10]As before, the second highest non-negative ironed virtual valuation is interpreted to be 0 if there are fewer than two bidders whose ironed virtual valuation is non-negative.

[11]The arrows in the notation denote "rounding down" for the infimum and "rounding up" for the supremum.

that is, for the first $i$ coordinates, we "round up" the inverse ironed virtual valuation to its supremum, and for the remaining $n-i$, we round it down to its infimum. Then it is obvious that $\chi(z_0) \leq \chi(z_1) \leq \cdots \leq \chi(z_n)$. Let $i_x$ be the index for which $\chi(z_{i_x-1}) \leq x \leq \chi(z_{i_x})$. We define $\overleftarrow{\mathbf{r}}(\nu_x) = z_{i_x-1}$ and $\overrightarrow{\mathbf{r}}(\nu_x) = z_{i_x}$.

Our final solution will be to pick one of the pricings $\overleftarrow{\mathbf{r}}(\nu_x)$ and $\overrightarrow{\mathbf{r}}(\nu_x)$ with $x = \max(x_0, 1/2)$, where $x_0$ is defined as follows: let $\mathbf{p}^0$ be an arbitrary pricing with $\bar\phi_{p_i{}^0} = 0$ for all $i$, and let $x_0 = \chi(\mathbf{p}^0)$. We will show that one of these pricings $\overleftarrow{\mathbf{r}}(\nu_x)$ and $\overrightarrow{\mathbf{r}}(\nu_x)$ is a 3-approximation to the revenue of $\mathcal{M}$.

**A better accounting**

We start with an accounting trick. $\mathcal{R}^\mathbf{p}$ is a complicated thing to calculate because when more than one item are priced below the consumer's values we must break ties in favor of the the item that gives the consumer the highest utility (i.e., with the maximum difference between the consumer's value and the item's prices). Our analysis in Section 3 used a crude lower bound $\mathcal{R}^\mathbf{p}$ in the case of two items are priced below value: the revenue is at least $\min_i p_i$, which if $\mathbf{p} = \mathbf{r}(\nu)$, is at least $\nu$ (by Fact 4). In this section we phrase our bounds more precisely in terms of this lower bound.

**Definition 6 ($\mathcal{Q}^\mathbf{p}$)** *For any $\nu$, let $\mathbf{p}$ be any pricing satisfing $\bar\phi_i(p_i) = \nu$ for all $i$. The revenue lower bound $\mathcal{Q}^\mathbf{p}$ is the expectation of the random variable $Z$, for $\mathbf{v}$ distributed as $\mathbf{F}$, that is,*

$$Z = \begin{cases} 0 & \text{if } v_i \leq p_i \text{ for all } i, \\ p_i & \text{if } v_j > p_j \text{ iff } j = i, \\ \nu & \text{otherwise.} \end{cases}$$

**Fact 25** *For any $\nu$ and $\mathbf{p}$ satisfing $\bar\phi_i(p_i) = \nu$, $\mathcal{R}^\mathbf{p} \geq \mathcal{Q}^\mathbf{p}$.*

If we replaced $\mathcal{R}^\mathbf{p}$ with $\mathcal{Q}^\mathbf{p}$ in Section 3 all statements would remain correct. Thus,

**Theorem 26** *With regular $\mathbf{F}$ and $\nu = \max(0, \nu_{1/2})$, the pricing $\mathbf{p} = \mathbf{r}(\nu)$ satisfies $\mathcal{Q}^\mathbf{p} \geq \mathcal{R}^\mathcal{M}/3$.*

In this section we show that for the $\mathbf{p}$ we choose that $\mathcal{Q}^\mathbf{p} \geq \mathcal{R}^\mathcal{M}/3$. Fact 25, then, implies our desired result.

**A regular analogy**

In order to analyze the non-regular setting, we will show that there is a regular instance with properties very close to that of our non-regular instance, so that some of the analysis from Section 3 carries over to the non-regular setting. In particular, for a given irregular distribution $F$ there exists a regular distribution $\bar F$ satisfying the property that the virtual valuation function for a value distributed from $\bar F$ is equal to the ironed virtual valuation function for a value distributed from $F$. This justifies a notational overlap where we let $\bar\phi(\cdot)$ represent the virtual valuation function with respect to $\bar F$. Recall that if $R_F(\alpha)$ denotes the expected revenue corresponding to distribution $F$ as a function of the probability of allocation, $\bar R_F(\alpha)$ denotes its concave envelope, or "ironed" revenue[12]. We define $\bar F$ in such a way that its expected revenue is $R_{\bar F}(\alpha) = \bar R_F(\alpha)$ for all $\alpha$.

Construct $\bar F$ from $\bar R$ as follows:

1. Define function $g(\alpha) = \bar R(\alpha)/(1-\alpha)$.

2. We show below that $g(\alpha)$ is strictly increasing for $\alpha \in [0, 1)$, so $g^{-1}(\cdot)$ is well defined.

3. Define $\bar F(v) = g^{-1}(v)$.

**Lemma 27** *In the above construction $g(\alpha)$ is strictly increasing for $\alpha \in [0, 1)$.*

---
[12]We drop the subscripts whenever there is no ambiguity.

*Proof:* Upon differentiating $g(\alpha)$ we get:

$$\begin{aligned}\frac{d}{d\alpha}g(\alpha) &= \frac{d}{d\alpha}\frac{\bar{R}(\alpha)}{1-\alpha} \\ &= \frac{\bar{R}(\alpha)}{(1-\alpha)^2} - \frac{\bar{\phi}(F^{-1}(\alpha))}{1-\alpha} \\ &= \frac{\bar{R}(\alpha) - (1-\alpha)\bar{\phi}(F^{-1}(\alpha))}{(1-\alpha)^2} \\ &> 0\end{aligned}$$

where the last inequality follows by observing for $\alpha < 1$:

$$(1-\alpha)\bar{\phi}(F^{-1}(\alpha)) < \int_\alpha^1 \bar{\phi}(F^{-1}(\alpha))d\alpha = \bar{R}(\alpha)$$

□

Note that, by definition, $R_{\bar{F}}(\alpha) = (1-\alpha)\bar{F}^{-1}(\alpha) = (1-\alpha)g(\alpha)$, for all $\alpha$. Therefore we have,

**Fact 28** *For all $\alpha$, $R_{\bar{F}}(\alpha) = \bar{R}_F(\alpha)$, and $\phi_{\bar{F}}(\bar{F}^{-1}(\alpha)) = \bar{\phi}_F(F^{-1}(\alpha))$.*

This fact immediately implies that $R_{\bar{F}}(\cdot)$ is concave, and therefore, $\phi_{\bar{F}}(\cdot)$ is non-decreasing.

**Lemma 29** *$\bar{F}$ is regular.*

**Approximate pricing**

For all $i$, define $\bar{F}_i$ given $R_{F_i}$ as above, and let $\bar{\mathbf{F}} = \bar{F}_1 \times \cdots \times \bar{F}_n$. Let $\mathcal{R}_{\bar{\mathbf{F}}}^\mathcal{M}$ be the revenue of Myerson's mechanism for the instance given by joint distribution $\bar{\mathbf{F}}$.

**Lemma 30** $\mathcal{R}_{\bar{\mathbf{F}}}^\mathcal{M} = \mathcal{R}_{\mathbf{F}}^\mathcal{M}$

*Proof:* Recall that in the regular case, the revenue of Myerson's mechanism is exactly equal to its expected virtual surplus. That is,

$$\mathcal{R}_{\bar{\mathbf{F}}}^\mathcal{M} = \int_{\bar{\mathbf{F}}} \max\{0, \max_i \phi_{\bar{F}_i}(\bar{F}_i^{-1}(\alpha))\}d\alpha$$

Likewise, in the non-regular case, the revenue of Myerson's mechanism is exactly equal to its expected ironed virtual surplus. Therefore,

$$\mathcal{R}_{\mathbf{F}}^\mathcal{M} = \int_{\mathbf{F}} \max\{0, \max_i \bar{\phi}_{F_i}(F_i^{-1}(\alpha))\}d\alpha$$

The lemma now follows from Fact 28. □

Let $x = \max\{x_0, 1/2\}$. Recall that we pick the best of the two pricings, $\overleftarrow{\mathbf{p}} = \overleftarrow{\mathbf{r}}(\nu_x)$, and $\overrightarrow{\mathbf{p}} = \overrightarrow{\mathbf{r}}(\nu_x)$. In addition, it will be useful for the analysis to consider the price vector $\mathbf{p}$ with $p_j = \overleftarrow{p_j} = \overrightarrow{p_j}$ for all $j \neq i_x$, and $p_{i_x} \in [\overleftarrow{p_{i_x}}, \overrightarrow{p_{i_x}}]$ defined such that $\chi(\mathbf{p}) = x$. The following lemma is just a restatement of Theorem 26 for the joint distribution $\bar{\mathbf{F}}$.

**Lemma 31** $\mathcal{Q}_{\bar{\mathbf{F}}}^\mathbf{p} \geq \mathcal{R}_{\bar{\mathbf{F}}}^\mathcal{M}/3$

On the other hand, the next lemma shows that at least one of $\overleftarrow{\mathbf{p}}$ and $\overrightarrow{\mathbf{p}}$ is superior to $\mathbf{p}$ in terms of the revenue lower bound $\mathcal{Q}$.

**Lemma 32** *Given any regular product distribution $\bar{\mathbf{F}}$, and a price vector $\mathbf{p}$ satisfying $\phi_i(p_i) = \nu$ for all $i$, let $\mathbf{p}'$ and $\mathbf{p}''$ be defined such that $p_j = p_j' = p_j''$ for all $j \neq i$, $p_i' = \inf\{v : \phi_i(v) = \nu\}$, and $p_i'' = \sup\{v : \phi_i(v) = \nu\}$. Then, $\max\{\mathcal{Q}_{\bar{\mathbf{F}}}^{\mathbf{p}'}, \mathcal{Q}_{\bar{\mathbf{F}}}^{\mathbf{p}''}\} \geq \mathcal{Q}_{\bar{\mathbf{F}}}^\mathbf{p}$.*

*Proof:* Let $y_0$ be the probability that for all indices $j \neq i$, $v_j < p_j$, $y_1$ be the probability that for exactly one index $j \neq i$, $v_j \geq p_j$, and $y_2$ be the probability that for at least two indices $j \neq i$, $v_j \geq p_j$. We can write $\mathcal{Q}^{\mathbf{p}}$ as

$$\mathcal{Q}^{\mathbf{p}} = p_i q_i y_0 + w(1-q_i)y_1 + \nu(q_i y_1 + y_2)$$

where $w$ is the expected revenue of the pricing $\mathbf{p}$ conditioned on the events that $v_j \geq p_j$ for exactly one index $j \neq i$, and $v_i < p_i$. Then,

$$\mathcal{Q}^{\mathbf{p}} = A + B(p_i q_i - C q_i)$$

where $A$, $B$ and $C$ are some constants independent of $p_i$ (but depend on $p_j$ and $q_j$ for $j \neq i$).

Consider maximizing the function $\mathcal{Q}^{\mathbf{p}}$ as a function of $p_i$ over the range $[p_i', p_i'']$. Differentiating with respect to $p_i$, we get

$$\tfrac{d}{dp_i}\mathcal{Q}^{\mathbf{p}} = B(C - \phi_i(p_i))f_i(p_i) = B(C - \nu)f_i(p_i)$$

Since $f_i(p_i)$ is always positive, and $C$ and $\nu$ are constant over $p_i \in [p_i', p_i'']$, this derivative is either always positive or always negative over the range $[p_i', p_i'']$. In the former case, $\mathcal{Q}^{\mathbf{p}}$ is maximized at $p_i''$, and in the latter case it is maximized at $p_i'$. $\square$

**Corollary 33** $\max\{\mathcal{Q}_{\mathbf{F}}^{\overleftarrow{\mathbf{p}}}, \mathcal{Q}_{\mathbf{F}}^{\overrightarrow{\mathbf{p}}}\} \geq \mathcal{Q}_{\mathbf{F}}^{\mathbf{p}}$.

Finally, we show that at prices $\overleftarrow{\mathbf{p}}$ and $\overrightarrow{\mathbf{p}}$, the revenue lower bound $\mathcal{Q}$ under the distribution $\mathbf{F}$ is equal to that under $\bar{\mathbf{F}}$.

**Lemma 34** $\mathcal{Q}_{\mathbf{F}}^{\overleftarrow{\mathbf{p}}} = \mathcal{Q}_{\bar{\mathbf{F}}}^{\overleftarrow{\mathbf{p}}}$, and likewise, $\mathcal{Q}_{\mathbf{F}}^{\overrightarrow{\mathbf{p}}} = \mathcal{Q}_{\bar{\mathbf{F}}}^{\overrightarrow{\mathbf{p}}}$

*Proof:* We will prove this lemma for the prices $\overleftarrow{\mathbf{p}}$. The proof for $\overrightarrow{\mathbf{p}}$ is analogous. Note that for all $i$, $\phi_{F_i}(\overleftarrow{p_i}) = \bar{\phi}_{F_i}(\overleftarrow{p_i})$ by the definition of $\overleftarrow{\mathbf{p}}$. Therefore, $\phi_{F_i}(\overleftarrow{p_i}) = \phi_{\bar{F}_i}(\overleftarrow{p_i})$. Then the events, $v_i \leq \overleftarrow{p_i}$ for all $i$, and $v_j > \overleftarrow{p_j}$ for some $j$, have equal probabilities under $\mathbf{F}$ and $\bar{\mathbf{F}}$. The lemma now follows from the definition of $\mathcal{Q}^{\mathbf{p}}$. $\square$

We can now use the above lemmas to prove the main result of this section:

**Theorem 35** Let $x = \max\{x_0, 1/2\}$, $\overleftarrow{\mathbf{p}} = \overleftarrow{\mathbf{r}}(\nu_x)$, and, $\overrightarrow{\mathbf{p}} = \overrightarrow{\mathbf{r}}(\nu_x)$. Then, $\max\{\mathcal{R}^{\overleftarrow{\mathbf{p}}}, \mathcal{R}^{\overrightarrow{\mathbf{p}}}\} \geq \mathcal{R}^{\mathcal{M}}/3$. That is, one of the two pricings $\overleftarrow{\mathbf{p}}$ and $\overrightarrow{\mathbf{p}}$ is a 3-approximation to the optimal pricing.

*Proof:* Let $\mathbf{p}$ be as above: $p_j = \overleftarrow{p_j} = \overrightarrow{p_j}$ for all $j \neq i_x$, and $p_{i_x} \in [\overleftarrow{p_{i_x}}, \overrightarrow{p_{i_x}}]$ defined such that $\chi(\mathbf{p}) = x$. Then, we have the following sequence of inequalities.

$$\begin{aligned}
\max\{\mathcal{R}^{\overleftarrow{\mathbf{p}}}, \mathcal{R}^{\overrightarrow{\mathbf{p}}}\} &\geq \max\{\mathcal{Q}_{\mathbf{F}}^{\overleftarrow{\mathbf{p}}}, \mathcal{Q}_{\mathbf{F}}^{\overrightarrow{\mathbf{p}}}\} && \text{Fact 25}\\
&= \max\{\mathcal{Q}_{\bar{\mathbf{F}}}^{\overleftarrow{\mathbf{p}}}, \mathcal{Q}_{\bar{\mathbf{F}}}^{\overrightarrow{\mathbf{p}}}\} && \text{Lemma 34}\\
&\geq \mathcal{Q}_{\bar{\mathbf{F}}}^{\mathbf{p}} && \text{Lemma 33}\\
&\geq \mathcal{R}_{\bar{\mathbf{F}}}^{\mathcal{M}}/3 && \text{Lemma 31}\\
&= \mathcal{R}_{\mathbf{F}}^{\mathcal{M}}/3 && \text{Lemma 30}
\end{aligned}$$

$\square$

We note that Theorem 35 only gives a characterization of an approximately optimal pricing in the non-regular case, and not a polynomial-time approximation algorithm. We leave open the problem of designing a polynomial time algorithm for this case (in particular, a polynomial time algorithm for computing ironed virtual valuations), noting that for the case when each of the distributions $F_i$ is discrete and explicitly specified, a simple algorithm for computing ironed virtual valuations has been given by Elkind (2007), and this implies a polynomial-time approximation algorithm for the non-regular case with discrete explicit distributions.

# 5 A polynomial-time approximation algorithm

We now describe how to implement our algorithm for the regular case in the two computational models described in Section 2. Notice that our analysis has reduced the multi-dimensional optimization problem (approximately optimal pricing) to a single dimensional optimization problem (optimal uniform virtual pricing). The remaining challenge we face will be in inverting virtual valuation functions, which we can only do approximately. Our final $3 + \epsilon$ approximation is guaranteed with a $1 - o(1)$ probability. In Section 5.3 we provide a simpler and faster algorithm that guarantees a $6 + \epsilon$ approximation with probability 1.

## 5.1 The discrete case

Implementation in the discrete explicit model is straightforward. Although we have focused on continuous distributions in Sections 3 and 4, we remark that virtual valuations and their inverses for discrete distributions can be defined and computed in much the same way as for continuous distributions (See, e.g., Elkind (2007)). The straightforward algorithm for the discrete case computes virtual valuations of all possible values for each item. During this process it keeps track of $F_i(\phi_i^{-1}(\nu))$. It then picks the least non-negative $\nu$ that satisfies $\chi(\mathbf{r}(\nu)) \leq 1/2$ and outputs $\mathbf{p} = \mathbf{r}(\nu)$. The running time of each step of the algorithm is at most linear in $n$ and the sizes of the supports.

This process may choose $\mathbf{p}$ with $\chi(\mathbf{p}) < 1/2$. To apply the analysis of Section 3, notice that choosing any $\nu' > \nu$ will have $\chi(\mathbf{r}(\nu')) > 1/2$. This discontinuity is due to the fact that we have a discrete distribution. Fortunately for our analysis, when $p_i = v_i$ the consumer has zero utility for buying the item and zero utility for buying nothing. In the setting we are in, there exists a tie-breaking rule for $\mathbf{p}$ such that $\chi(\mathbf{p}\text{ with tie-breaking}) = 1/2$. It is easy to see that this tie-breaking rule is not an optimal tie-breaking rule; recall that our implicit tie-breaking rule that favors the most expensive item is optimal. Thus, the revenue of pricing $\mathbf{p}$ (with our implicit tie-breaking rule) is at least the revenue of $\mathbf{p}$ with the new tie-breaking rule. Since $\mathbf{p}$ with the new tie-breaking rule is a 3-approximation to Myerson, so is $\mathbf{p}$ (with the implicit rule).

## 5.2 The continuous case

In the continuous case, recall that we assume the distributions are specified via oracles for sampling and for evaluating the cumulative distribution function $F_i$ and the density function $f_i$. This case is challenging because the oracle model does not allow for exact computation of inverse virtual valuations; it only allows these quantities to be approximated. We construct an algorithm based on the analysis of Section 3. This approach is based on computing virtual valuations $\phi_i(v_i)$ for valuations $v_i$ that are powers of $(1+\epsilon)$, collecting a set of candidate pricings such that our previous analysis guarantees there is a good pricing among the collection, and then sampling from the distribution to output the pricing with the best empirical performance.[13] The main technical innovation in this section is a proof that shows if we can coordinate-wise approximate a pricing, we can approximate it revenue-wise as well.

### 5.2.1 From coordinate-wise to revenue-wise approximations

Central to our construction is a result which shows that if we can find a price point, $\mathbf{p}''$, that is close (e.g., in $L_\infty$ distance) to some desired price point, $\mathbf{p}$, then we can find a new point $\mathbf{p}'$ from $\mathbf{p}''$ that has revenue close to $\mathbf{p}$. This result confirms the suspicion that if a pricing can be approximated (point-wise) then so can its revenue. This result is a corollary of a lemma due to Nisan (See Balcan et al. (2005)).[14]

**Lemma 36** *For $\epsilon \in (0,1)$, let $\mathbf{p}$ and $\mathbf{p}'$ be pricings that satisfy $p_i' \in [1-\epsilon, 1+\epsilon^2 - \epsilon] \, p_i$ for all $i$. Then $\mathcal{R}^{\mathbf{p}'} \geq (1-2\epsilon)\mathcal{R}^{\mathbf{p}}$.*

---

[13]Because of the sampling step, we can only guarantee that with high probability the price we output is good.

[14]Though we do not discuss the details, Lemma 36 and Corollary 37 hold true even in settings where the consumers have general combinatorial preferences.

*Proof:* Consider any valuation vector $\mathbf{v}$, and let $i$ be the index that maximizes $v_i - p_i$. In other words, when prices are given by $\mathbf{p}$ and a consumer has values $\mathbf{v}$, the consumer buys item $i$. On the other hand, let $j$ be the index that maximizes $v_j - p_j'$. That is, when the prices are given by $\mathbf{p}'$, the same consumer buys item $j$ instead of $i$. The lemma follows from the claim that

$$p_j' \geq (1 - 2\epsilon)p_i.$$

To prove this claim, we first observe that $v_i - p_i \geq v_j - p_j$ and $v_j - p_j' \geq v_i - p_i'$. Rearranging terms and adding the two we get $p_i - p_i' \leq p_j - p_j'$. Finally, using $p_i' \leq (1 + \epsilon^2 - \epsilon)p_i$ and $p_j' \geq (1 - \epsilon)p_j$, we get

$$(\epsilon - \epsilon^2)p_i. \leq \epsilon p_j$$

The claim now follows by dividing by $\epsilon$ and again using the fact that $p_j' \geq (1 - \epsilon)p_j$. □

**Corollary 37** *For $\epsilon \in (0, 1)$, let $\mathbf{p}$, $\mathbf{p}'$, and $\mathbf{p}''$ be pricings that satisfy, for all $i$, $p_i' = (1 + \epsilon^2 - \epsilon)p_i''$ and $p_i'' \in [1 - \epsilon^2, 1] \, p_i$. Then $\mathcal{R}^{\mathbf{p}'} \geq (1 - 2\epsilon)\mathcal{R}^{\mathbf{p}}$.*

*Proof:* Simply verify that $(1 - \epsilon^2)(1 + \epsilon^2 - \epsilon) \geq 1 - \epsilon$ so the conditions of Lemma 36 apply to $\mathbf{p}'$ and $\mathbf{p}$. □

### 5.2.2 Our algorithm

Let $M = (\max_i h_i)/(\min_i \ell_i)$. Our algorithm will run in time polynomial in $n$ and $M$. For each item $i$, we consider the set $L_i$ of values that are powers of $\gamma = 1/(1 - \epsilon)$ for some $\epsilon > 0$ in the range $[\ell_i, h_i]$. Note that $|L_i| = O(\log_\gamma \frac{h_i}{\ell_i}) = O(\log_\gamma M)$.

**Definition 7 (Approximate Uniform Virtual Price Algorithm)** *Parameterized by constants $\delta$ and $\epsilon$ in $(0, 1)$, the approximate uniform virtual price algorithm proceeds as follows:*

1. *Evaluate $\chi(\mathbf{r}(0))$. If $\chi(\mathbf{r}(0)) \geq 1/2$ then output pricing $\mathbf{p} = \mathbf{r}(0)$ and stop.*

2. *For each $i$ and each $v \in L_i$, compute $\phi_i(v)$ using the oracles for $F_i$ and $f_i$ and store these.*

3. *For any $\nu$, let $\mathbf{r}'(\nu) = (r_1'(\nu), \ldots, r_n'(\nu))$ where $r_i'(\nu)$ is the largest value in $L_i$ whose virtual value is at most $\nu$.*

4. *Let $L' = \{\nu \mid \nu \in \bigcup_i L_i \text{ and } \chi(\mathbf{r}'(\nu)) \leq 1/2\}$.*

5. *Let $P = \{\mathbf{p} \mid p_i = (1 + \epsilon^2 - \epsilon)r_i'(\nu') \text{ and } \nu' \in L'\}$.*

6. *Let $S$ be an i.i.d. sampling of $\frac{4M^2}{\epsilon^2} \log \frac{\delta}{2|P|}$ valuation profiles from $\mathbf{F}$.*

7. *Output $\mathbf{p} = \operatorname{argmax}_{\mathbf{p}' \in P} \mathcal{R}^{\mathbf{p}'}(S)$.*

We first note that one of the pricings $\mathbf{p} \in P$ is near the pricing $\mathbf{r}(\nu_{1/2})$ that our analysis of Section 3 suggests.

**Lemma 38** *If $\chi(\mathbf{r}(0)) < 1/2$ then there is a $\nu' \in L'$, as defined in the approximate uniform virtual price algorithm, that satisfies*

$$(1 - \epsilon)r_i(\nu_{1/2}) \leq r_i'(\nu') \leq r_i(\nu_{1/2})$$

*for all $i$.*

*Proof:* First we note that $\mathbf{r}'(\nu_{1/2})$ satisfies $(1 - \epsilon)r_i(\nu_{1/2}) \leq r_i'(\nu_{1/2}) \leq r_i(\nu_{1/2})$ for all $i$. Then we show that $\mathbf{r}'(\nu_{1/2}) = \mathbf{r}'(\nu')$ for some $\nu' \in L'$.

1. For any $\nu$, $\mathbf{r}'(\nu)$ satisfies $(1-\epsilon)r_i(\nu) \leq r'_i(\nu) \leq r_i(\nu)$ for all $i$. This follows from the fact that $L_i$ contains values that are powers of $1/(1-\epsilon)$ and $\phi_i^{-1}(\nu)$ falls between two such powers. $r'_i(\nu)$ is the lower of these two powers.

2. Consider $\mathbf{r}'(\nu_{1/2})$. Notice that in $\mathbf{r}'(\nu_{1/2})$ each item $i$ has a price that corresponds to a virtual value of at most $\nu_{1/2}$. Let $\nu'$ be the highest of these virtual valuations. Then $\mathbf{r}'(\nu') = \mathbf{r}'(\nu_{1/2})$. Furthermore, since $\nu' \leq \nu_{1/2}$, $\chi(\mathbf{r}'(\nu')) \geq 1/2$. We conclude that $\nu'$ is in $L'$.

□

The following theorem follows directly from Corollary 37, Lemma 38, and Lemma 40 (below).

**Theorem 39** *For any $\delta$ and $\epsilon$ in $(0,1)$, with probability $1-\delta$ the approximate uniform virtual price algorithm gives a $(3+O(\epsilon))$-approximation to BUPP in time polynomial in $n$, $M$ and $1/\epsilon$.*

### 5.2.3 A sampling lemma

The next building block in our approach to approximating the pricing suggested by Section 3 is a statement that shows that sampling from the distribution will allow us to accurately compare several pricings to determine which has higher expected revenue. For completeness we give the full proof, though this result follows directly from standard approaches.

For a particular consumer with valuation vector $\mathbf{v} = (v_1, \ldots, v_n)$, let $\mathcal{R}^{\mathbf{p}}(\mathbf{v})$ be the actual revenue when $\mathbf{p}$ is offered them and they choose their favorite item. Let $\mathcal{R}^{\mathbf{p}}(S)$ be the average revenue of $\mathbf{p}$ when offered to each of the consumers in a set $S$. Our next lemma shows that for a large enough sample consumers $S$ drawn from our distribution, $\mathcal{R}^{\mathbf{p}}(S)$ is a good approximation to $\mathcal{R}^{\mathbf{p}}$.

**Lemma 40** *For any set $P$ of pricings with $\mathbf{p} \in P$ satisfying $\chi(\mathbf{p}) \geq 1/2$ and any sample $S$ of $|S| \geq \frac{4M^2}{\epsilon^2} \log \frac{\delta}{2|P|}$ independent draws from $\mathbf{F}$,*

$$\mathbf{Pr}[\forall \mathbf{p} \in P, \ |\mathcal{R}^{\mathbf{p}}(S) - \mathcal{R}^{\mathbf{p}}| \geq \epsilon \mathcal{R}^{\mathbf{p}}] \leq \delta,$$

*where $M = \max_{\mathbf{p},i} p_i / \min_{\mathbf{p},i} p_i$.*

*Proof:* Of course, for $\mathbf{v}$ drawn from $\mathbf{F}$, $\mathbf{E}[\mathcal{R}^{\mathbf{p}}(\mathbf{v})] = \mathcal{R}^{\mathbf{p}}$. We also have $\mathcal{R}^{\mathbf{p}}(\mathbf{v}) \leq \max_i p_i$ for all $\mathbf{v}$, and $\mathcal{R}^{\mathbf{p}} \geq \frac{1}{2} \min_i p_i$ because $\chi(\mathbf{p}) \geq \frac{1}{2}$. Apply the Chernoff bound to get,

$$\mathbf{Pr}[|\mathcal{R}^{\mathbf{p}}(S) - \mathcal{R}^{\mathbf{p}}| \geq \epsilon \mathcal{R}^{\mathbf{p}}] \leq 2 \exp\left(-\frac{(\epsilon \mathcal{R}^{\mathbf{p}})^2}{\sum_{j \in S} (\max_i p_i / |S|)^2}\right)$$

$$\leq 2 \exp\left(-\frac{\epsilon^2 |S|}{4M^2}\right)$$

$$\leq \frac{\delta}{|P|}.$$

The last step follows from the assumption that $|S| \geq \frac{4M^2}{\epsilon^2} \log \frac{\delta}{2|P|}$. Now take the union bound over all $\mathbf{p} \in P$ to prove the lemma. □

## 5.3 A simple $(6+\epsilon)$-approximation

The $(3+\epsilon)$-approximation algorithm, though it follows from a direct approach, is rather cumbersome. In this section we show how to obtain a conceptually and algorithmically simple algorithm. The algorithm here gives a single pricing that is close to optimal and does not require sampling the distribution.

The algorithm is based on the observation that the Vickrey auction with appropriate non-anonymous reservation prices (henceforth, "Vickrey with reserve prices") is a 2-approximation to the optimal single-item auction. A similar construction to that in Section 3 can then be applied to derive a pricing that mimics

Vickrey with reserve prices (and is a 3-approximation to it). This construction will only require us to compute the inverse virtual valuation of zero, i.e., the optimal sale price, for each of the distributions. We assume that computing the optimal sale price for each distribution $F_i$ is possible. (Approximations can also be used in this step, if necessary.) Once again we assume that the distributions $F_i$ are regular.

### 5.3.1 Vickrey with reserve prices

The standard interpretation of Myerson's main result is that the optimal auction is Vickrey with an appropriate reservation price. This, of course, is only true under regularity and when the bidders' valuations are identically distributed. Myerson's main result, that maximizing welfare is equivalent to maximizing (ironed) virtual surplus, solves the single-item optimal auction problem (and more general single-parameter agent problems) even when regularity does not hold and when the agent valuations are not identically distributed. A natural question that, to our knowledge, has not been previously answered, is measuring the extent to which Vickrey with reservation prices approximates the optimal single-item auction when valuations are not identically distributed. We show that, with a natural choice of reservation prices, it is a 2-approximation.

**Definition 8** *The* Vickrey auction with reserve prices $\mathbf{p} = (p_1, \ldots, p_n)$, $\mathcal{V}_{\mathbf{p}}$, *sells an item to the bidder $i$ with bid $b_i \geq p_i$ that has the highest bid, if such a bidder exists. The payment that this winning bidder makes is the maximum of their reservation price $p_i$ and the bid of the highest other bidder $j$ that bids at least their reservation price, $p_j$. When the bids are distributed from $\mathbf{F}$ we call $\mathbf{p} = \mathbf{r}(0)$ the optimal reservation prices.*

**Lemma 41** *The Vickrey auction with the optimal reservation prices, $\mathcal{V}_{\mathbf{r}(0)}$, is a 2-approximation to the optimal single-item auction.*

*Proof:* This lemma holds both in the regular and the non-regular case. We first prove it in the regular case. We are drawing valuations $\mathbf{v}$ from distribution $\mathbf{F}$ and comparing the expected profit of the optimal single-item auction, $\mathcal{M}$ (with profit $\mathcal{R}^{\mathcal{M}}$), and the Vickrey auction with optimal reserve prices, $\mathcal{V}_{\mathbf{r}(0)}$ (with profit $\mathcal{R}^{\mathcal{V}_{\mathbf{r}(0)}}$). Let $i$ be the winner of $\mathcal{V}_{\mathbf{r}(0)}$ and $j$ be the winner of $\mathcal{M}$ (we let $i = j = 0$ represent the case that there are no winners). Notice that $i = j = 0$ if and only if all virtual valuations are less than zero; and otherwise, $v_i \geq v_j$ (since $\mathcal{V}_{\mathbf{r}(0)}$ allocates the highest valued bidder with non-negative virtual valuation).

This proof uses the fact that the expected sum of payments in an auction is equal to the expected virtual surplus (Theorem 1). Thus,

$$\mathcal{R}^{\mathcal{M}} = \mathbf{E}[\phi_j(v_j)]$$
$$= \mathbf{E}[\phi_j(v_j) \mid i = j]\mathbf{Pr}[i = j] + \mathbf{E}[\phi_j(v_j) \mid i \neq j]\mathbf{Pr}[i \neq j]$$

Both parts of this last expression can be bounded from above by $\mathcal{R}^{\mathcal{V}_{\mathbf{r}(0)}}$. For the first part,

$$\mathbf{E}[\phi_j(v_j) \mid i = j]\mathbf{Pr}[i = j] = \mathbf{E}[\phi_i(v_i) \mid i = j]\mathbf{Pr}[i = j]$$
$$\leq \mathbf{E}[\phi_i(v_i)]$$
$$= \mathcal{R}^{\mathcal{V}_{\mathbf{r}(0)}}.$$

The inequality above follows from the non-negativity of virtual surplus. We bound the second part by $\mathcal{R}^{\mathcal{V}_{\mathbf{r}(0)}}$ using the fact that $\phi_j(v_j) \leq v_j$ (Fact 4) and the fact that the payment made in $\mathcal{V}_{\mathbf{r}(0)}$ is at least $v_j$ as follows.

$$\mathbf{E}[\phi_j(v_j) \mid i \neq j]\mathbf{Pr}[i \neq j] \leq \mathbf{E}[v_j \mid i \neq j]\mathbf{Pr}[i \neq j]$$
$$\leq \mathbf{E}[i\text{'s payment} \mid i \neq j]\mathbf{Pr}[i \neq j]$$
$$\leq \mathbf{E}[i\text{'s payment}]$$
$$= \mathcal{R}^{\mathcal{V}_{\mathbf{r}(0)}}.$$

Above, the last inequality comes from the fact that $i$'s payment is always non-negative. Thus,

$$\mathcal{R}^{\mathcal{M}} \leq 2\mathcal{R}^{\mathcal{V}_{\mathbf{r}(0)}}.$$

The same proof holds in the non-regular case as well, upon replacing the virtual surplus and virtual values above by ironed virtual surplus and ironed virtual values respectively. Once again we use the fact that $\mathcal{V}_{\mathbf{r}(0)}$ always has a non-negative ironed virtual surplus (although it may sometimes have a negative virtual surplus in the non-regular case). □

### 5.3.2 Approximating Vickrey by Pricing

This section follows almost identically to Section 3 where we showed that there is a pricing based on the Myerson's optimal auction that approximates its profit. Here we show that there is a pricing based on the Vickrey auction with optimal reservation prices that approximates its profit.

The pricing we consider will be $\mathbf{p} = (p_1, \ldots, p_n)$ with $p_i = \max(r_i(0), v)$ for $v$ chosen such that the probability that there is a winner is exactly $1/2$ (if no such $v$ exists then $p_i = r_i(0)$). The first step of our analysis will be to relate $\mathcal{R}^{\mathcal{V}_{\mathbf{r}(0)}}$ to $\mathcal{R}^{\mathcal{V}_{\mathbf{p}}}$ which is analogous to our relating $\mathcal{R}^{\mathcal{M}}$ to $\mathcal{R}^{\mathcal{M}_\nu}$.

**Lemma 42** *Under regularity, with $v$ and $\mathbf{p}$ with $p_i = \max(v, r_i(0))$ satisfying $\chi(\mathbf{p}) = 1/2$,*

$$\mathcal{R}^{\mathbf{p}} \geq \mathcal{R}^{\mathcal{V}_{\mathbf{r}(0)}}/3.$$

*Proof:* Consider $\mathcal{V}_{\mathbf{p}}$ and recall that $\mathcal{V}_{\mathbf{p}}$ allocates if and only if the pricing $\mathbf{p}$ does. Let $x = \chi(\mathbf{p}) = \chi(\mathcal{V}_{\mathbf{p}})$ be the probability that neither allocates. Let $q_i$ be the probability that $v_i \geq p_i$, i.e., $q_i = 1 - F_i(p_i)$. We proceed in steps.

1. Observe that when $\mathcal{V}_{\mathbf{r}(0)}$ allocates but $\mathcal{V}_{\mathbf{p}}$ does not its sale price is less than $v$. Thus,

$$\mathcal{R}^{\mathcal{V}_{\mathbf{r}(0)}} \leq \mathcal{R}^{\mathcal{V}_{\mathbf{p}}} + vx.$$

2. Recall that the pricing $\mathbf{p}$ allocates if and only if $\mathcal{V}_{\mathbf{p}}$ allocates and when it allocates its revenue is at least $v$. Thus,

$$v(1-x) \leq \mathcal{R}^{\mathbf{p}}.$$

3. Lemma 8 shows that for any $\mathbf{p}$,

$$\mathcal{R}^{\mathbf{p}} \geq x \sum_i p_i q_i.$$

4. Lemma 9 applied to $\mathcal{V}_{\mathbf{p}}$ shows,

$$\mathcal{R}^{\mathcal{V}_{\mathbf{p}}} \leq \sum_i p_i q_i.$$

5. Combine the above inequalities and recall that $x = 1/2$. Thus,

$$\mathcal{R}^{\mathbf{p}} \geq \mathcal{R}^{\mathcal{V}_{\mathbf{r}(0)}}/3.$$

□

Combining the lemma above with Lemma 13 (which shows that $\mathcal{R}^{\mathbf{r}(0)} \geq \mathcal{R}^{\mathcal{M}}$ when $\chi(\mathbf{r}(0)) > 1/2$) and Corollary 2 (which shows that $\mathcal{R}^{\mathcal{M}} \geq \mathcal{R}^{\mathcal{V}_{\mathbf{r}(0)}}$) we obtain the main theorem of this section.

**Theorem 43** *If $\chi(\mathbf{r}(0)) > 1/2$ let $\mathbf{p} = \mathbf{r}(0)$; otherwise, let $\mathbf{p}$ with $p_i = \max(p_i, v)$ satisfy $\chi(\mathbf{p}) = 1/2$. Then assuming regularity,*

$$\mathcal{R}^{\mathbf{p}} \geq \mathcal{R}^{\mathcal{V}_{\mathbf{r}(0)}}/3.$$

### 5.3.3 Computing p

First, we assume that we are given $r_i = \phi_i^{-1}(0)$. (Once again, Corollary 37 implies that $(1 - \epsilon^2)$-approximations to these worsen our approximation factor by a multiplicative $1 + O(\epsilon)$.) These are the optimal sale prices for each item if we were selling them alone. Now we need to compute a $\nu$ such that **p** with $p_i = \max(r_i, \nu)$ the probability no items is sold (i.e., $x$ from the above section), is (approximately) $1/2$. This can be done by binary search to arbitrary accuracy. Notice that if we do not have $x = 1/2$ exactly, plugging $x = 1/2 - \epsilon$ into the above theorem gives us $\mathcal{R}^\mathbf{p} \geq \mathcal{R}^{\mathcal{V}_{\mathbf{r}(0)}}/(3 + \epsilon)$ (and using Lemma 41, $\mathcal{R}^\mathbf{p} \geq \mathcal{R}^\mathcal{M}/(6 + \epsilon)$).

**Theorem 44** *There is a polynomial time algorithm that gives a $(6 + \epsilon)$-approximation to BUPP in the regular case.*

## 6 Conclusions

Several interesting questions related to BUPP still remain open:

- Is the Bayesian unit-demand pricing problem with independently distributed values NP-hard to solve optimally? There is some evidence that this problem is indeed hard. For example, one can construct two-item instances with extremely simple distributions (e.g., a uniform distribution over some range), where the optimal price is irrational.

- Is our characterization tight? Can one construct an example where the revenue of Myerson's auction is indeed three times the revenue of the optimal pricing?

  It is worth noting that there is a simple example in which the revenue of the pricing defined by our virtual valuation technique falls short of the optimal pricing by a factor of nearly 2, even in the i.i.d. case. Suppose that for each $i$, the distribution of $v_i$ is given by

  $$\begin{align} \mathbf{Pr}[v_i = n] &= \tfrac{1}{n^2} \\ \mathbf{Pr}[v_i = 1] &= 1 - \tfrac{1}{n^2}. \end{align}$$

  The optimal pricing sets $p_1 = 1$ and $p_i = n$ for all $i > 1$. This achieves a revenue of $2 - o(1)$. However, for every $\nu$ the pricing which sets $p_i = \phi^{-1}(\nu)$ achieves a revenue of at most 1. (In this example the revenue of Myerson's auction is nearly equal to 2, so the example does not prove any separation between the revenue of Myerson's auction and that of the optimal pricing.)

- Extending this work to accommodate combinatorial consumers seems tricky. An optimal pricing in that case may offer bundles at prices higher or lower than the sum of the prices of individual items in the bundle.

- Finally, a more general selling mechanism in the unit-demand case may offer lotteries to consumers. A lottery is a distribution over single items, sold at a price (typically) lower than the prices for the individual items. The revenue of the optimal collection of lotteries is not always bounded above by the revenue of Myerson's auction. Furthermore, when values are correlated, the revenue of the optimal single-item pricing can be an exponential factor smaller than the revenue of the optimal collection of lotteries.

## Acknowledgments

We thank Nina Balcan for several useful discussions.

# References

Aggarwal, G., Feder, T., Motwani, R., Zhu, A., 2004. Algorithms for multi-product pricing. In: Proceedings of ICALP.

Aggarwal, G., Hartline, J., 2006. Knapsack auctions. In: Proceedings of the 17th Annual ACM-SIAM Symposium on Discrete Algorithms.

Armstrong, M., 2006. Recent Developments in the Economics of Price Discrimination. Vol. 2 of Advances in Economics and Econometrics: Theory and Applications. Cambridge University Press, pp. 97–141.

Balcan, M.-F., Blum, A., Hartline, J., Mansour, Y., 2005. Mechanism design via machine learning. In: Proc. of the 46th IEEE Symp. on Foundations of Computer Science.

Balcan, N., Blum, A., 2006. Approximation algorithms and online mechanisms for item pricing. In: Proc. 8th ACM Conf. on Electronic Commerce.

Briest, P., Krysta, P., 2007. Buying cheap is expensive: Hardness of non-parametric multi-product pricing. In: Proceedings of the 18th ACM-SIAM Symposium on Discrete Algorithms.

Bulow, J., Roberts, J., 1989. The simple economics of optimal auctions. The Journal of Political Economy 97, 1060–90.

Demaine, E., Feige, U., Hajiaghayi, M., Salavatipour, M., 2006. Combination can be hard: Approximability of the unique coverage problem. In: Proceedings of the 17th ACM-SIAM Symposium on Discrete Algorithms.

Elkind, E., 2007. Designing and learning optimal finite support auctions. In: Proceedings of the 18th ACM-SIAM Symposium on Discrete Algorithms.

Guruswami, V., Hartline, J., Karlin, A., Kempe, D., Kenyon, C., McSherry, F., 2005. On profit-maximizing envy-free pricing. In: Proceedings of the 16th ACM-SIAM Symposium on Discrete Algorithms.

Hartline, J., Koltun, V., 2005. Near-optimal pricing in near-linear time. In: Proceedings of the 9th Workshop on Algorithms and Data Structures. Springer-Verlag.

Myerson, R., 1981. Optimal auction design. Mathematics of Operations Research 6, 58–73.

Rockafellar, R. T., 1970. Convex Analysis. Princeton University Press.

Stole, L. A., 2007. Price Discrimination and Competition. Vol. 3 of Handbook of Industrial Organization. Elsevier, Ch. 34, pp. 2221–2299.